# Spatially resolved stress measurements in materials with polarization-sensitive optical coherence tomography: image acquisition and processing aspects

(short running title: Spatially resolved stress measurements with PS-OCT)

B. Heise[1,2], K. Wiesauer[1], E. Götzinger[3], M. Pircher[3], C.K. Hitzenberger[3], R. Engelke[4], G. Ahrens[4], G. Grützner[4] and D. Stifter[5]

[1]Upper Austrian Research GmbH, Hafenstrasse 47-51, 4020 Linz, Austria

[2]FLLL – Department of Knowledge-Based Mathematical Systems, Johannes Kepler University Linz, Altenbergerstrasse 69, 4040 Linz, Austria

[3]Center for Biomedical Engineering and Physics, Medical University of Vienna, Währingerstr. 13, 1090 Wien, Austria

[4]micro resist technology GmbH, Köpenicker Str. 325, 12555 Berlin, Germany

[5]ZONA - Center for Surface- and Nanoanalytics, Johannes Kepler University Linz, Altenbergerstrasse 69, 4040 Linz, Austria
[*]david.stifter@jku.at

**ABSTRACT**: We demonstrate that polarization-sensitive optical coherence tomography (PS-OCT) is suitable to map the stress distribution within materials in a contactless and non-destructive way. In contrast to transmission photoelasticity measurements the samples do not have to be transparent but can be of scattering nature. Denoising and analysis of fringe patterns in single PS-OCT retardation images are demonstrated to deliver the basis for a quantitative whole-field evaluation of the internal stress state of samples under investigation.

**Introduction**

Photoelasticity is currently the method of choice to experimentally determine the stress state within a specimen [1,2]. The technique works in general in transmission and involves the detection of stress or load induced birefringence in a sample by evaluating a (series of) recorded fringe pattern(s). A drawback of this method is related to the fact that the specimen has to be transparent or that a model of the sample made from a transparent (polymer) material has to be used.

Optical coherence tomography (OCT) is a measurement technique recently developed for biomedical diagnostics applications [3]: OCT provides in a contactless and non-destructive way high-resolution cross-sectional images of the internal structure of turbid and scattering tissue and materials. Besides the original OCT method, a multitude of instrumental extensions and modifications has been presented in the last years and is still a main theme in OCT research, involving developments like polarization-sensitive OCT (PS-OCT) [4]. PS-OCT delivers in addition to the depth-resolved intensity images, which are related to the internal backscattering structure of the sample, the state of polarization of the light in the sample.

Besides the vast amount of applications of OCT and its derivates in medical diagnostics alternative applications outside the biomedical field are gaining momentum, as summarized in a recent review [5]. In this context it has been shown for PS-OCT, that anisotropies and internal strain/stress distribution can be determined in semi-transparent and scattering polymer and composite structures [6-9]. In this paper, we use for the first time PS-OCT to optimize etched polymer structures and apply, also for the first time and to the best of our knowledge, 2D demodulation techniques on single PS-OCT images to quantitatively determine the internal stress state of samples.

**Experimental set-up**
For OCT a broadband light-source operating in the near-infrared and an interferometer is used to determine position of scattering sites and reflecting interfaces in a sample, as depicted in Figure 1a): in general (for the so-called time-domain OCT), interference bursts are observed by moving the reference mirror. The envelope of the interference signal represents a single, one-dimensional reflectivity profile from the sample (termed A-scan). By performing A-scans at adjacent sites on the sample, cross-sectional images are acquired (B-scan).

By adding polarization optics, the OCT set-up can be made polarization sensitive. We follow an approach as schematically depicted in Figure 1b) and based on a concept presented in ref. [10]: the linearly polarized light from the light source is directed through broadband quarter waveplates situated in the interferometer arms. The waveplate in the reference arm is oriented in such a way that the light coming from this arm is linearly polarized with an angle of 45° with respect to the polarizing axes of the polarizing beamsplitter of the detection unit which consists of two single detectors, one for each polarization direction. The waveplate in the sample arm leads to an illumination of the sample with circularly polarized light. Any birefringence present in the sample yields in general elliptically polarized light reaching the polarization sensitive detection unit. With this approach the sample reflectivity R, the magnitude of the optical retardation δ (phase lag of one polarization direction with respect to the orthogonal one, induced by optical anisotropy in the plane perpendicular to the depth axis) and the orientation of the optical axis φ can simultaneously be measured as a function of the depth z [10]:

$$R(z) \sim (A_1(z))^2 + (A_2(z))^2 \qquad (1)$$

$$\delta(z) = \arctan(A_1(z)/A_2(z)) \qquad (2)$$

$$\phi = (180° - (\Phi_2 - \Phi_1))/2 \qquad (3)$$

with $A_i$ and $\Phi_i$ (i=1,2) being the amplitude and phase, respectively, of the observed oscillations of the interference signals on the individual detectors 1 and 2.

The set-ups used for the experiments presented in this paper include at first a PS-OCT system operating at a center wavelength of 1550 nm and exhibiting a depth resolution around 15 µm in typical polymer materials (n~1.5). The second PS-OCT used is operating at 800 nm center wavelength and shows ultrahigh-resolution capabilities (~2 µm depth resolution). In addition, this second PS-OCT set-up is capable of measuring so-called en-face scans (or C-scans), which are area scans taken at a fixed depth within the sample (scan parallel to the sample surface, perpendicular to the impinging light onto the sample). A detailed description of the set-ups and their characteristics can be found in refs. [6,9].

**Results and Discussion**

Cross-sectional and en-face PS-OCT imaging

For a better understanding and demonstration, we present at first PS-OCT images taken from a scattering polymer sample (LDPE, low density polyethylene) with a thickness of 1 mm. In the upper row of Figure 2 cross-sectional reflectivity images are shown for the sample, once in its original state (left) and once bent (right). The images were taken with the 1550 nm system with the light beam impinging on the sample surface from top (sample was not cut for cross-sectional imaging). It is worth mentioning that the full cross-sectional area is successfully imaged throughout the whole specimen thickness and that the backside of the polymer sample is clearly visible, although the sample is not transparent but of highly scattering nature. No influence of the bending procedure on the internal structure, which is dominated by a distinct speckle pattern, can be observed. As to the bent surface, one has to keep in mind, that the light is refracted at the curved air/polymer interface and that the images shown represent the optical path the light travelled, which can result in distortions between imaged structure and real geometry. In order to correct for this effect, either a flat surface can be prepared (e.g. with index matching gel on the sample surface) or by numerically correcting the images, as demonstrated e.g. with a ray-tracing procedure in [11]. As to the current depicted reflectivity images in Figure 2, no specific features, like microparticles, are visible in the interior and the speckle pattern shows no correlation between the unloaded and loaded state, ruling out in this case strain determination by correlation techniques like OCT elastography [12]. In contrast to the reflectivity images, the corresponding retardation images (bottom row) show the change of the birefringence state in the plane perpendicular to the depth axis within the sample induced by the bending procedure. The images are gray-scale encoded: a transition from black to white represents a phase lag of $\pi/2$ of the slow polarization direction with respect to the orthogonal, fast one (optical retardation). A further increase of the retardation leads the signal to move back towards black values (retardation value of $\pi$) and continues then to white ($3\pi/2$),… (signal mirrored at multiples of $\pi/2$). A high frequency of stripes consequently corresponds to a highly strained (anisotropic) sample structure, as also shown by us in [6,7].

From the demonstrational case presented in Figure 2 and from our previous work [6-9] it is evident that crucial information can be obtained with PS-OCT from the internal strain/stress state of a sample. A real application is reported in Figure 3. In this case we use en-face PS-OCT to evaluate LIGA (German acronym for Lithographie-Galvanisierung-Abformung) photo-resist structures fabricated in cooperation with the German BESSY synchrotron radiation source by X-ray depth lithography and serving as moulds for miniature/micro gear-wheels (wheel design by Micromotion GmbH, Germany). The structures are etched in thick SU-8 photo-resist layers (~1 mm thickness) deposited on gold-coated silicon wafers. The width of the structures can be as small as 30 μm, resulting in high aspect ratio trenches. Beside the determination of detrimental particles situated in the etched mould trenches the residual stress within the developed photo-resist, causing deformations and even cracks, is of interest before the final (and expensive) galvanization step is made. En-face PS-OCT images taken at the photo-resist/wafer interface can deliver both, information on geometry (dimensions, defects, cracks and particles) and information on residual stress (similar as reflection photoelasticity), as principally shown in ref. [9] and now in detail in this work: for the first time we used PS-OCT for the verification of the optimization of the photolithographic patterning process, as depicted in Figure 3.

The three images in the two rows of this figure show the reflectivity- (left), the retardation- (middle) and optical axis orientation images (right) of two structures etched in the thick SU8 photo-resist layer obtained with our en-face scanning PS-OCT working at 800 nm. The retardation- and also the optical axis orientation images are gray-scale encoded (retardation values analogue to those in Figure 2). The gray-values of the optical orientation images give the direction of the fast optical axis perpendicular to the incident light beam (0°-180°). The sample in the upper row demonstrates an initial status of lithographic process settings with high exposure dose. Increased stress is detected especially between the two outer ring trenches of the two wheels as indicated by the bright areas in the retardation image. The stress additionally caused a crack (indicated by the arrow in the upper intensity image). By en-face PS-OCT, a fact experienced in lithographic practice could be now substantiated, namely that a lower exposure dose results in a lower stress level in the resist material: stress and crack formation is greatly reduced at lower exposure as registered in a contact-free and non-destructive way with our en-face PS-OCT (bottom row). This way, en-face PS-OCT is a promising method for the verification of objective quality criteria in the photolithographic patterning process.

From the above presented examples and applications it is apparent that relevant qualitative information can be obtained on the strain state in samples by means of PS-OCT. However, mostly quantitative and spatially resolved data, especially on the internal stress distribution, is required. In case of cross-sectional PS-OCT images, the retardation is given in a cumulative way over the depth. To obtain the actual birefringence value at a certain depth position, the slope of the cumulative and unwrapped retardation fringe pattern needs to be determined [9]. If the stress-birefringence response of the respective material is known for the actual imaging wavelength, actual stress values can be given for each position within the sample. A procedure to determine the stress-birefringence response with PS-OCT is presented e.g. in ref. [7], with the stress-optical coefficient describing the dependence in the linear regime of the stress-birefringence dependence. To date, unwrapping of retardation values has been performed on the respective retardation A-scans to determine the slope [7,13] or in the case of ref. [14] on Stokes Vector A-scan profiles to yield a 2D stress map. However, the sequential 1D unwrapping procedure of A-scans has been used on relative simple stripe structures which are a result of a retardation monotonically increasing in depth (i.e. no change from tensile to compressive stress). For more complex stress situations, causing e.g. ring-like structures (closed fringes) as depicted in the retardation image of the right column of Figure 2 (or even better visible in the retardation image of Figure 5), a 2D image processing and evaluation approach has to be chosen. Therefore, we demonstrate in the following for the first time phase unwrapping of whole PS-OCT cross-sections (B-scans) based on a single image unwrapping algorithm.

PS-OCT image processing: denoising and phase unwrapping

Image processing of fringe patterns to recover quantitative deformation information is already a central task in speckle interferometry (SI) [15]. Processing of several phase shifted images is a standard routine in SI, although already single image algorithms have recently been developed [16]. We extend the latter technique to process our PS-OCT retardation images, as at first demonstrated in Figure 4 on a simple retardation fringe pattern obtained with our 1550 nm PS-OCT from a linearly stressed polymer sample (1 mm thick polymer bar made from LDPE, strained in load cell). The exact procedure consists of the following steps:

1) Denoising:
Cross-sectional retardation fringe images $I(x,z)$ of scattering materials exhibit a distinct speckle pattern, as seen in Figure 4a. Due to the low axial resolution of the used PS-OCT system (15 µm), the speckle noise is in this case much more severe than in standard SI images. With conventional denoising algorithms like wavelet based and median filtering methods we obtained only insufficient results. Therefore, we apply at first an anisotropic diffusion technique, in particular an algorithm based on coherence enhancing diffusion CED [17,18] for denoising and enhancement of the fringe structure, with the result as depicted in Figure 4b (fringe values normalized to values between -1 and 1).

2) Demodulation:
A 2D-quadrature demodulation [19] is performed on the enhanced and normalized retardation fringe pattern, since with our chosen PS-OCT approach conventional phase shift techniques can not be applied as usually done in interferometry for robustness. We exploit the recently developed technique for single fringe pattern analysis as realized e.g. in (holographic) interferometry or conventional photoelasticity [16,20] and test these algorithms for the analysis of our single PS-OCT retardation image.

At first, the quadrature image $I_Q(x,z)$ of the smoothed fringes $I_S(x,z)$ is computed (Figure 4c) by applying a 2D Hilbert transform, which can be realized by a spiral phase filter in the Fourier domain [21]. Additionally, an estimation of the fringe orientation is required [22]. For this estimation step we use a 2D energy operator based method as described in refs. [23,24]. The orientational unwrapping is then performed on the basis of a quality map similar to ref. [25] with the quality criteria adapted by us to be dependent on the angular gradient and signal-to-noise ratio. It is also worth noting, that in contrast to interferometric fringe patterns we can not assume harmonic functions for our retardation images (instead: triangular-shaped retardation functions in the A-scans, see e.g. Figure 2a in [10]). However, the influence of the deviation of the signal from a harmonic one can be neglected for the Hilbert transform as we have verified on simulated fringe patterns.

The wrapped retardation phase $\varphi_W(x,z)$ (Figure 4d) can be determined from the argument of the complex extension $\varphi_W(x,z) = arg\,[I_S(x,z) + i\,I_Q(x,z)]$, taking a periodicity of $\pi$ for the retardation into account, in contrast to the usual $2\pi$ periodicity for interferometry. A FFT- based 2D-unwrapping method [26,27] was finally performed to determine the unwrapped retardation $\varphi_{UW}(x,z)$, as depicted in Figure 4e.

3) Differentiation and stress mapping
In the last step, the smoothed and unwrapped retardation image is numerically differentiated in depth direction z to obtain the local birefringence value [9]. A least square optimization or regularization technique is used for suppressing high frequency noise during numerical differentiation. The actual stress distribution is finally determined by taking the stress/birefringence response of the actual material into account (Figure 4f). For the LDPE sample in Figure 4, the dependence of the birefringence on the stress $\sigma$ has already been determined in PS-OCT calibration experiments performed on this material [7]: in these experiments the LDPE sample bar was successively strained and the respective stress and birefringence values were obtained with a load cell and the 1550 nm PS-OCT set-up. A manual evaluation of single A-scans with respect to an average birefringence value of the strained sample is consistent with the now automatically determined unwrapped data. The before measured stress/birefringence data points from [7] have now been fitted with an exponential function and this function was used for the stress determination from the birefringence map.

The above described procedure is performed on a more complicated retardation pattern with partly closed fringes, as shown in the single retardation image of Figure 5 obtained with the 1550 nm PS-OCT set-up. A polymer bar, made from the same LDPE material as the sample in Figure 4, was clamped at both ends, which were moved towards each other, resulting in a mirrored double-S-like structure with areas of tensile strain on top of the resulting main bow and at the backside of the sample close to the clamped ends. The images in Figure 5 show a region around the top bow with areas of tensile and compressive strain, with the original retardation pattern in Figure 5a, the smoothed fringes in Figure 5b, the corresponding quadrature image in 5c, the wrapped and unwrapped phase image in 5d and 5e and the finally computed stress image in 5f, giving the difference between principal stresses in a plane perpendicular to the light beam. For the stress image, one has to keep in mind that the stress direction (tensile or compressive) has to be known at least in a single point in order to determine the direction in the whole 2D-image. In the case of Figure 5, the original retardation image is rather noisy, which could lead to unwrapped sub-areas changing sign, however, the single image algorithm correctly reflects in this case the expected stress geometry (tensile stress on top of the bow and at the backside of the sample close to the left side).

In the case of en-face PS-OCT images, the above described 2D unwrapping approach allows to determine a complete lateral stress map (with stress values averaged over the imaging depth), as shown in Figure 6 on a photo-resist gear-wheel structure. A region of interest has been taken, which corresponds to the narrowing located between two neighboring wheel structures and which was monitored in the experiments for process optimization presented above in Figure 3 with the 800 nm en-face PS-OCT set-up. For the calculation of the unwrapped image the denoising step could be omitted, since the original en-face retardation scan *I(x,y)* (Figure 6b) exhibits no speckle noise (smooth resist-wafer interface). The final stress image was obtained by dividing the resulting birefringence values with a stress-optical coefficient of $2.6 \times 10^{-4}$ $MPa^{-1}$, determined in ref. [28] for a wavelength of 635 nm. For a precise measurement of the stress, the stress-optical coefficient around wavelengths of 800 nm needs at first to be determined, but the current error is estimated to be less than 1% due to the flat slope of the dispersion curve of the refractive index of the SU-8 material in this wavelength region. A selected profile in Figure 6e) shows that at the center of the narrowing stress values up to 2.5 MPa are reached.

**Conclusions and Outlook**
We have demonstrated that PS-OCT works in reflection and is capable of delivering spatially resolved information on the internal birefringence/stress state in a plane perpendicular to the depth axis of samples, which can be – in contrast to standard photoelasticity – also of highly scattering nature. In combination with complementary phase-sensitive and interferometric methods which exhibit sensitivity along the depth axis [29,30], or by performing PS-OCT scans with varying incident angles, a full sample characterization is within reach. For the PS-OCT retardation images a 2D image evaluation procedure has been developed, based on algorithms taken from single fringe image analysis in conventional interferometry and photoelasticity and was tested on a simple retardation image with known birefringence/stress response. From the presented advanced applications and by taking the novel developments and recent trends in the field of advanced OCT imaging into account it can be expected that PS-OCT will play an increasingly important role in the field of photomechanics. In this view, new light sources providing a shorter coherence length in the 1500 nm range will deliver images with a finer speckle structure by simultaneously ensuring a higher penetration depth (compared to 800 nm sources, see e.g. [5] for a penetration depth study in polymers). Furthermore, PS-OCT systems realized in the ultrafast and highly sensitive Fourier-domain technique [31,32] will allow to rapidly capture complete 3D PS-OCT data sets and put dynamically performed strain/stress testing with fully automated image analysis within reach.

**ACKNOWLEDGEMENTS**
This work has been financially supported by the Austrian Science Fund FWF (Projects L126-N08 and P19751-N20). We thank Joachim Weickert for his support in image enhancement by CED methods and Andreas Jablonski for discussions on image denoising aspects.

**FIGURES**

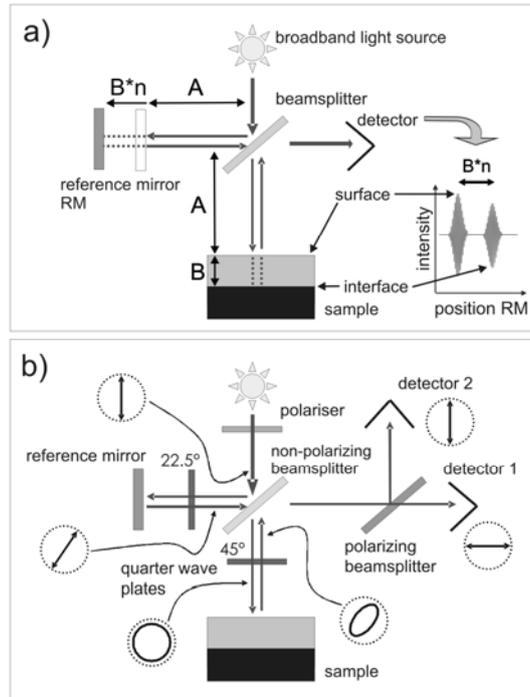

Figure 1: a) Schematic sketch of an OCT set-up with movable reference mirror (A, B: distances, n: refractive index of sample layer). b) Extension of OCT towards PS-OCT by adding polarizer, wave plates and polarization sensitive detection unit (pictograms in dotted circles indicate state of polarization).

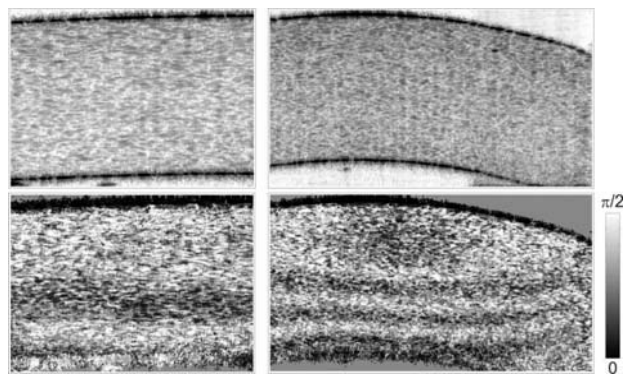

Figure 2: Cross-sectional PS-OCT images of a turbid polymer sample (thickness 1 mm), once in its original state (left column) and once bent (right column). The top row gives the intensity values and the bottom row the corresponding retardation images (gray scale encoded, mirrored at multiple $\pi/2$ values).

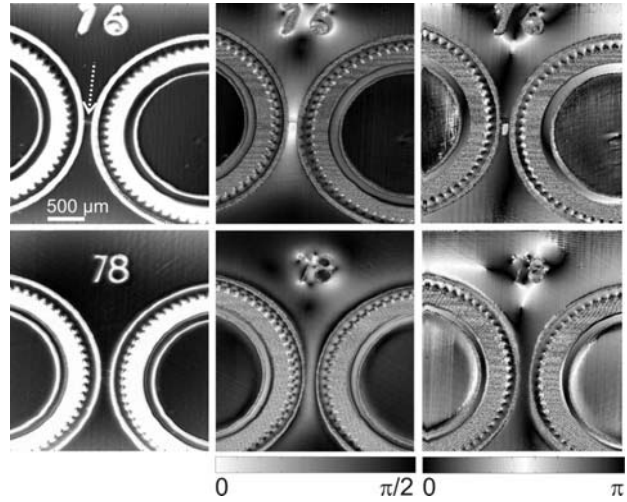

Figure 3: En-face PS-OCT images taken from photo-resist moulds for miniature gear-wheels made by X-ray lithography (top row: highly strained resist mould, bottom row: structure with lower strain resulting from processing with lower exposure dose). Left: reflectivity images parallel to the surface at a depth of 1 mm (arrow indicates crack due to high stress. Middle: corresponding retardation scans indicating regions of strain (gray scale encoded: 0-$\pi/2$ optical retardation). Right: orientation of the fast optical axis (gray scale encoded: 0-$\pi$) mapping the direction of the strain field.

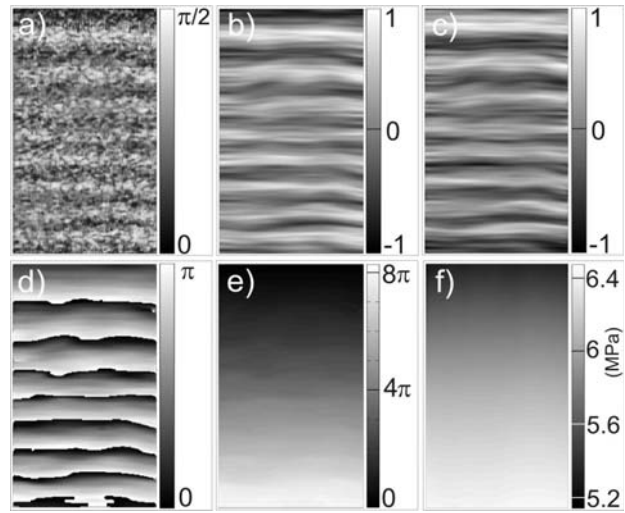

Figure 4: Single image phase unwrapping procedure for cross-sectional PS-OCT images (example of a simple stripe pattern). a) Original PS-OCT image $I(x,z)$ exhibiting distinct speckle structure, b) denoised image $I_S(x,z)$ (values normalized from -1 to 1), c) quadrature image $I_Q(x,z)$ of b), d) resulting wrapped phase image $\varphi_W(x,z)$, e) unwrapped phase image $\varphi_{UW}(x,z)$, f) corresponding stress image $\sigma(x,z)$.

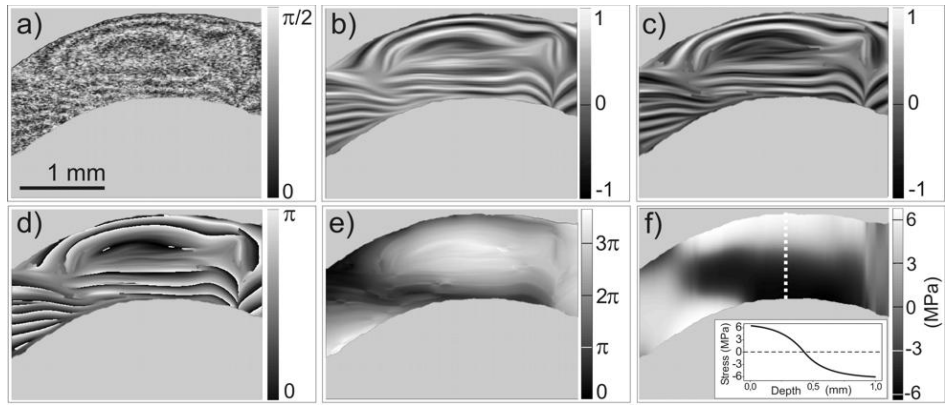

Figure 5: Unwrapping procedure for a complex cross-sectional PS-OCT image (bent polymer structure). a) Original PS-OCT image, b) denoised image, c) quadrature image, d) wrapped phase image, e) unwrapped phase image, f) stress image (inset: stress profile along indicated dotted line).

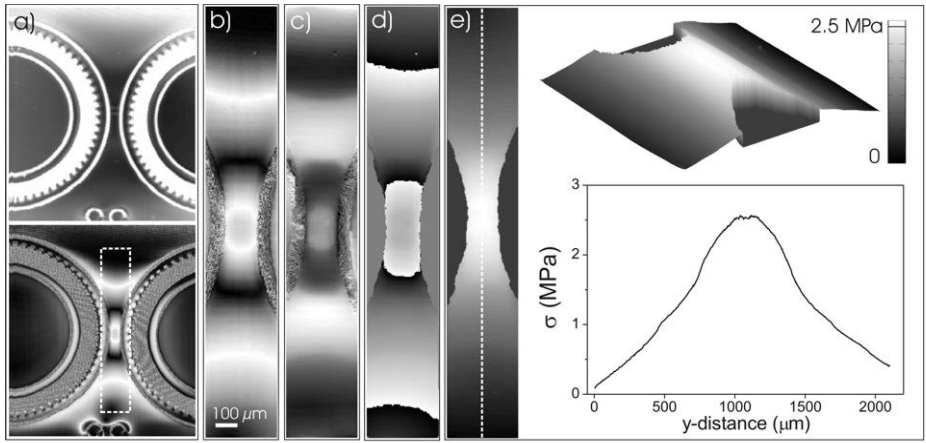

Figure 6: Unwrapping procedure for en-face PS-OCT images. a) Original intensity (top) and retardation (bottom) images. b) enlarged view of area indicated in retardation image of a), c) quadrature image, d) wrapped phase image, e) stress image (plane and 3D view, as well as profile along indicated dotted line). Gray-values of a)-d) are used in analogy to those of the respective images in Figure 4 and 5.

10